\documentclass[letterpaper, 10 pt, conference]{ieeeconf} 
\IEEEoverridecommandlockouts                              
\usepackage{amsmath}
\usepackage{mathtools}
\usepackage{amsfonts}
\usepackage{amssymb}
\usepackage{algorithm} 
\usepackage{algpseudocode}
\usepackage{color}
\usepackage{dblfloatfix}
\usepackage{balance}
\usepackage[normalem]{ulem}

\newtheorem{definition}{Definition}
\newtheorem{lemma}{Lemma}
\newtheorem{remark}{\bf Remark}

\newtheorem{theorem}{Theorem}
\newtheorem{corollary}{Corollary}

\newcommand{\diagentry}[1]{\mathmakebox[1.8em]{#1}}
\newcommand{\xddots}{%
  \raise 4pt \hbox {.}
  \mkern 10mu
  \raise 1pt \hbox {.}
  \mkern 10mu
  \raise -2pt \hbox {.}
}

\newcommand{\xddotswide}{%
  \raise 5pt \hbox {.}
  \mkern 10mu
  \raise 2pt \hbox {.}
  \mkern 10mu
  \raise -1pt \hbox {.}
}

\title{\LARGE \bf Sequential Synthesis of Distributed Controllers for \\Cascade Interconnected Systems}
\author{Etika Agarwal\textsuperscript{*}, S. Sivaranjani\textsuperscript{*}, Vijay Gupta and Panos Antsaklis
\thanks{The authors are with the Department of Electrical Engineering, University of Notre Dame, Notre Dame, IN. \{eagarwal, sseethar, vgupta2, pantsakl\}@nd.edu. \newline
Funding: E. Agarwal and P. Antsaklis were funded by ARO under grant no. ARL W911NF-17-1-0072. The work of S. Sivaranjani
was supported in part by NSF grants CNS-1544724 and ECCS-1550016, and
that of V. Gupta by NSF grant CNS-1739295 and ARO grant W911NF-17-1-0072.\newline
\textsuperscript{*}These authors contributed equally to this work.
}
}

\begin{document}
\maketitle
\thispagestyle{empty}
\pagestyle{empty}

\begin{abstract}
We consider the problem of designing distributed controllers to ensure passivity of a large-scale interconnection of linear subsystems connected in a cascade topology. The control design process needs to be carried out at the subsystem-level with no direct knowledge of the dynamics of other subsystems in the interconnection. We present a distributed approach to solve this problem, where  subsystem-level controllers are locally designed in a sequence starting at one end of the cascade using only the dynamics of the particular subsystem, coupling with the immediately preceding subsystem and limited information from the preceding subsystem in the cascade to ensure passivity of the interconnected system up to that point. We demonstrate that this design framework also allows for new subsystems to be compositionally added to the interconnection without requiring redesign of the pre-existing controllers.
\end{abstract}

\section{Introduction}
Large-scale interconnected system architectures consisting of several dynamically coupled subsystems are increasingly being encountered in several infrastructure networks. For example, in power grids with high renewable energy penetration, it has been proposed that several small power sources and loads may be aggregated into clusters known as microgrids, which are interconnected to form the large-scale power network \cite{olivares2014trends}. Similarly, in large-scale transportation networks, vehicles with integrated communication can be operated in platoons to increase fuel efficiency, improve traffic congestion and enhance safety \cite{varaiya1993smart}\cite{sadraddini2017provably}. To guarantee stability, robustness and performance in such large-scale interconnected systems, distributed and decentralized control implementations 
that do not require that individual controllers have access to the states of all subsystems in the network, have been proposed to decrease the computational and communication costs. 
However, in typical distributed and decentralized control implementations, the design process itself is centralized, assuming knowledge of the dynamics, or even disturbances of other subsystems \cite{antonelli2013interconnected}-\nocite{wang1973stabilization}\nocite{lau1972decentralized}\nocite{aoki1971some}\nocite{bellman1974large}\nocite{davison1990decentralized}\nocite{vidyasagar1980decomposition}\nocite{farokhi2014decentralized}\cite{langbort2004distributed}. In large-scale infrastructure networks, it may be infeasible for each subsystem to have knowledge of the precise dynamics of other subsystems. Further, new subsystems may be added at a later stage or during operation, requiring a re-analysis of the interconnected system to ensure stability and performance, which may be intractable for large-scale interconnections.

In this context, distributed synthesis of controllers, where the control design for individual subsystems is carried out with limited knowledge of the dynamics of other subsystems in the network, is an important problem. Distributed synthesis of controllers has been explored in \cite{bakule1988decentralized}-\nocite{sezer1986nested}\nocite{sethi1998near}\nocite{zeilinger2013plug}\nocite{sivaranjani2017distributed}\cite{riverso2014plug} by either assuming, or designing controllers to impose, weak coupling between subsystems. However, such assumptions of weak dynamical coupling between subsystems may neither be desirable nor be practical in several physical scenarios. For example, dynamical coupling between subsystems is important in power sharing between microgrids in a power network, and weak coupling assumptions may only be made in the case of islanded operation \cite{tucci2016plug}. Furthermore, several such approaches rely on model predictive control schemes with control invariant set computations that may be computationally complex for large-scale interconnections \cite{zeilinger2013plug}\cite{riverso2014plug}. 

An alternative approach is offered by passivity and dissipativity-based control designs. Passive systems, under mild conditions, are known to possess useful properties such as stability \cite{haddad2008nonlinear} and demonstrate compositionality under feedback and parallel interconnection architectures, that is, negative feedback or parallel interconnections of passive systems are passive. This compositionality property can be helpful in distributing the control design process of large-scale systems when feedback and parallel interconnections are involved. Analysis of passivity properties of large-scale systems with star-shaped and cyclic symmetries is considered in \cite{wu2011passivity}\cite{ghanbari2016large}. Passivity of more general interconnection topologies of passive subsystems is discussed in \cite{arcak2016networks}-\nocite{vidyasagar1979new}\nocite{moylan1979tests}\cite{pota1993stability}, where information about the passivity property of all the subsystems is used to verify passivity of their interconnection. While these methods do not use any direct information about the subsystem dynamics , most still use a centralized procedure which requires passivity information of all subsystems for passivity verification of the interconnected system. Clearly, these approaches may not be scalable to large-scale systems and dynamically growing interconnections where subsystems may be added to the interconnection at a later time.

Therefore, it is useful to develop a theory for distributed synthesis of controllers using passivity-based tools that is applicable to general interconnection topologies and promotes compositionality for dynamically growing interconnections. As a first step towards this goal,  we consider the problem of distributed synthesis of local controllers for the special case of cascade interconnected linear systems. Passivity analysis and control design for series interconnected systems, a special case of cascade interconnections, was considered in \cite{yu2010passivity}; however, the design process was again centralized. In contrast, we propose a distributed procedure for passivity based control design of cascade interconnected systems.

We aim to design local controllers at the subsystem-level to ensure passivity of the overall interconnected system, while requiring that the design process at the subsystem-level only uses knowledge of the dynamics of the individual subsystem dynamics and information about its coupling with other subsystems. We solve this problem in three stages. 
\begin{itemize}
    \item First, we develop a \textit{sequential verification} approach to ascertain passivity of cascade interconnected systems. In this method, the passivity condition for the interconnected system is distributed into conditions on the passivity of individual subsystems and the coupling between them. Passivity verification for the interconnected system is carried out sequentially at the individual subsystem-level as follows. Starting from one end of the cascade, each subsystem utilizes its own dynamics and the coupling with its preceding neighbor, along with limited information in the form of a \textit{messenger matrix} communicated from the neighbor to verify the passivity of the interconnection up to that subsystem.
    \item Next, we use this sequential verification result to distribute the control design process. We propose a \textit{sequential control design} process, wherein beginning at one end of the cascade, each subsystem synthesizes a local controller using only the knowledge of its own dynamics and the messenger matrix received from its preceding neighbor to guarantee passivity of the entire interconnected system up to that point.
    \item Finally, we demonstrate that this design allows for new subsystems to be added `\textit{compositionally}' to the cascade, that is, without requiring redesign of the pre-existing controllers. Only the messenger matrix from the point of coupling, and the dynamics of the newly added subsystem are used to design the local controller for the new subsystem. 
    \end{itemize}




\vspace{0.2em}
\textit{Notation:} We denote the sets of real numbers, positive real numbers including zero, and $n$-dimensional real vectors by $\mathbb{R}$, $\mathbb{R}^{+}$ and $\mathbb{R}^{n}$ respectively. Define $\mathbb{N}_N=\{1,\ldots,N\}$, where $N$ is a natural number excluding zero. Given a block matrix $A  = \left[A_{i,j}\right]_{i\in\mathbb{N}_n,j\in\mathbb{N}_m}$, $A_{i,j}$ represents the $(i,j)$-th block, and $A'$ represents its transpose. Given matrices $A_{1},\ldots, A_{i}$, $\mbox{diag}(A_{1},\ldots,A_{i})$ represents a block-diagonal matrix with $A_{1},\ldots, A_{i}$ as its diagonal entries. A symmetric positive (semi-)definite matrix $P$ is represented as $P>0$ ($P\geq 0$). The identity matrix is denoted by $\mathbf{I}$, with dimensions clear from the context. Given sets $A$ and $B$, $A\backslash B$ represents the set of all elements of $A$ that are not in $B$. 

\section{System Dynamics and Problem Description}\label{sec:sys_dyn}
\begin{figure}[b]
    \centering
    \vspace{-1.4em}
    \includegraphics[scale=0.4,trim=0cm 0.5cm 0cm 0.1cm]{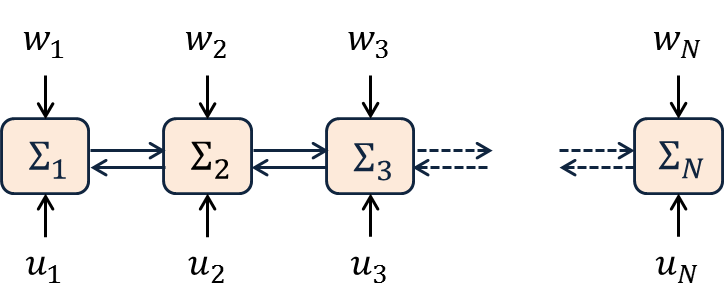}
    \caption{Cascade interconnected system $\Sigma$.}
    \label{fig:cascade}
  \vspace{-0.1em}
\end{figure}

Consider a cascade interconnected system $\Sigma$ as shown in Fig. \ref{fig:cascade}, comprised of $N$ subsystems $\Sigma_i$, $i \in \mathbb{N}_N$. The dynamics of the $i$-th subsystem $\Sigma_i$ is described by
\begin{subequations}
\begin{align}\label{linear system}
\dot{x}_i(t) & = A_i x_i(t) + B_i^{(1)}v_i(t) + B_i^{(2)} w_i(t) + B_i^{(3)} u_i(t),\nonumber \\
y_i(t) & = C_ix_i(t),\end{align} 
\vspace{-1.1em}
\begin{align}
v_i(t) & = \begin{cases} h_{i,i} x_i(t) + h_{i,i+1} x_{i+1}(t), & \hspace{-2em} \mbox{if } i = 1 \\
h_{i,i-1} x_{i-1}(t) + h_{i,i} x_i(t) + h_{i,i+1} x_{i+1}(t),\\ & \hspace{-5.2em}\mbox{if } i\in\mathbb{N}_{N-1} \backslash \{1\} \\
h_{i,i-1} x_{i-1}(t) + h_{i,i} x_i(t),  & \hspace{-2em}\mbox{if } i = N.
\end{cases}
\end{align}
\end{subequations}
where \(x_i(t) \in \mathbb{R}^{n_i} \), \(y_i(t) \in \mathbb{R}^{m_i}\),   \(v_i(t)\in \mathbb{R}^{p_i}\), \(w_i(t)\in\mathbb{R}^{m_i}\) and \(u_i(t)\in \mathbb{R}^{p_i}\) are the state, output, coupling input, disturbance and control input respectively. Define $x(t)$ to be the system state formed by stacking states $x_{i}(t)$ of all the $N$ subsystems. Similarly define $y(t)$, $v(t)$, $w(t)$, and $u(t)$ as being formed by stacking the outputs $y_{i}(t)$, coupling input $v_{i}(t)$, disturbance $w_{i}(t)$ and control input $u_{i}(t)$, $i \in \mathbb{N}_N$, respectively. The dynamics of the interconnected system $\Sigma$~is
\begin{equation}\label{interconnected_linear}
    \begin{aligned}
	\dot{x}(t) & = Ax(t) + B^{(1)}v(t) + B^{(2)}w(t) + B^{(3)}u(t)\\
	y(t) & = Cx(t)\\
	v(t) & = Hx(t) \\
	H & = \left[h_{i,j}\right]_{i,j\in\mathbb{N}_N}, \quad h_{i,j} = 0 \ \forall \vert i-j\vert>1
	\end{aligned}
	\end{equation}
where $A{=}\mbox{diag}(A_1, A_2, \dots, A_N)$, $B^{(j)}{=}\mbox{diag}(B^{(j)}_1, B^{(j)}_2,$ $ \dots,B^{(j)}_N), \, j\in \mathbb{N}_3$, and $C =\mbox{diag}(C_1, C_2, \dots, C_N)$.
In \eqref{interconnected_linear}, $H$ is the interconnection matrix, also referred to as \textit{coupling matrix}. Define the \textit{neighbor index set} of subsystem $\Sigma_i$, $i\in \mathbb{N}_N$ to be $\mathcal{I}_i=\{j: \vert i-j \vert < 2, j\in\mathbb{N}_N\}$. Also define $n=\sum\limits_{i=1}^{N}n_i$ and $m=\sum\limits_{i=1}^{N}m_i$. Note that the system model in this paper differs from typical cascade interconnection in literature in the sense that the dynamic coupling between subsystems is bidirectional. 
We now state some definitions that will be useful in this work. 
 \begin{definition}\cite{nijmeijer1993passive}\label{def:SSIP}
 A dynamical system \eqref{interconnected_linear} is said to be state-strictly passive (SP) from $w$ to $y$, with $u\equiv0$, if there exists $\epsilon>0$, and a positive definite function $V(x):\mathbb{R}^n\longrightarrow\mathbb{R}^+$, such that, for all $t>t_0\geq0$, $x(t_0)\in\mathbb{R}^n$ and $w(t)\in\mathbb{R}^m$
 \begin{equation}
     \int_{t_0}^{t} \big(w'(\tau)y(\tau)-\epsilon x'(\tau)x(\tau)\big)d\tau > V(x(t))-V(x(t_0))
 \end{equation}
 holds, where $x(t)$ is the state at time $t$ resulting from the initial condition $x(t_0)$.
 \end{definition}

We henceforth use the terms passivity and state-strict passivity (SP) interchangeably.
\begin{figure*}[!b]
	\newcounter{MYtempeqncnt}
    \setcounter{MYtempeqncnt}{\value{equation}}
    \setcounter{equation}{6}
    \vspace{0.5em}
    \small
    \hrulefill
    \begin{equation}\label{messenger_matrix}
    \begin{aligned} 
    \mathcal{M}_i &= \begin{cases} \mathcal{S}_1, & i =1  \\ \mathcal{S}_i - \mathcal{F}_i, & i \in \mathbb{N}_N \backslash \{1\}
        \end{cases} \\
    \hat{H}_{i,j} &= Q_iB_i^{(1)}H_{i,j}, \hspace{16.9em} i\in\mathbb{N}_N,\, j \in \mathcal{I}_i \\
    \mathcal{S}_i & = -(A'_i Q_i+Q_i A_i)-(\hat{H}_{i,i}+\hat{H}'_{i,i}) - \epsilon_i \mbox{I}, \hspace{5.3em} i \in \mathbb{N}_N
    \\\mathcal{F}_i &= (\hat{H}'_{i-1,i}+\hat{H}_{i,i-1})\mathcal{M}_{i-1}^{-1}(\hat{H}_{i-1,i}+\hat{H}'_{i,i-1}), \hspace{4em} i \in \mathbb{N}_N \backslash \{1\}
    \end{aligned}
    \end{equation}
\vspace{-1em}
\end{figure*}
\setcounter{equation}{3}
\normalsize

 The process of enforcing passivity of a system through a suitable control design is called \textit{passivation} and such a controller is referred to as \textit{passivating controller}. Typical techniques for design of passivating controllers scale poorly for large-scale interconnections. In this context, the objective of this paper is to introduce a distributed design framework to obtain passivating controllers for large-scale cascade interconnections of linear systems that scales well even when more subsystems are dynamically added to the interconnection. The term distributed design refers to the process of designing subsystem-level controllers locally at every subsystem with no direct information about the dynamics of other subsystems, while simultaneously guaranteeing state-strict passivity of the interconnected system. 
 Specifically, given the cascade interconnection $\Sigma$, we aim to 
\begin{enumerate}
	\item formulate a procedure for distributed verification of the passivity of \eqref{interconnected_linear} at the subsystem-level,
	\item propose a design technique to locally synthesize controllers at the subsystem-level, with no direct knowledge of the dynamics of other subsystems, to guarantee passivity of the interconnected linear system \eqref{interconnected_linear}, and 
	\item ensure compositionality of the control design, i.e., develop an algorithm to guarantee passivity of a dynamically growing interconnection, such that the addition of new subsystems does not require redesigning the pre-existing local controllers in the network.
\end{enumerate}


\section{Sequential Control Synthesis}\label{sec:seq}
In this section, we describe a sequential control design approach to synthesize subsystem-level controllers such that the interconnected system $\Sigma$ is passive. Beginning at one end of the cascade interconnection, subsystem controllers are locally designed in a sequence using only the model of the particular subsystem and limited information communicated from the preceding subsystem in the cascade. When a new subsystem is added to the cascade, only information from its neighboring subsystem is used to design a local controller for the new subsystem, while maintaining passivity of the interconnected system. 
\subsection{Sequential verification}
We begin by presenting a sequential technique for the verification of passivity of the interconnected system. We derive two properties of positive definite matrices that will be used to distribute the passivity verification. The proofs of the results in this section are collected in the Appendix.
\begin{lemma}\cite[Section 4.2]{golub2012matrix}\label{lemma:Cholesky}
A symmetric matrix $P$ is positive definite if and only if there exists a lower triangular matrix $L$ with positive diagonal entries such that $P = LL'$. Additionally, if such an $L$ exists then it is unique.
\end{lemma}

\begin{lemma}\label{lemma:Cholesky_tri_diagonal}
A symmetric block tri-diagonal matrix
\footnotesize
\begin{equation}\label{tri_diag_matrix}
    P = \begin{bmatrix} 
    \diagentry{\mathcal{P}_1} & \diagentry{\mathcal{R}_2} \\
    \diagentry{\mathcal{R}'_2} & \diagentry{\mathcal{P}_2} & \diagentry{\mathcal{R}_3} & & \text{\huge 0}\\
    &\diagentry{\mathcal{R}'_3} & \diagentry{\mathcal{P}_3} & \diagentry{\mathcal{R}_4}\\
    &&\diagentry{\xddots}& \diagentry{\xddots} & \diagentry{\xddots}\\
    &\text{\huge 0}&& \diagentry{\mathcal{R}'_{N-1}} & \diagentry{\mathcal{P}_{N-1}} & \diagentry{\mathcal{R}_N}\\
    &&&&\diagentry{\mathcal{R}'_N} & \diagentry{\mathcal{P}_N}\\
    \end{bmatrix},
\end{equation}\normalsize
where $\mathcal{P}_i$, $\mathcal{R}_i$, $i\in\mathbb{N}_N$ are block matrices of appropriate dimension, is positive definite if and only if
    \begin{equation}\label{seq_Cholesky_lemma}
    \begin{aligned}
    {M}_i & > 0,\hspace{8em}  \forall i \in \mathbb{N}_N,  \\
        {M}_i & =\begin{cases} \mathcal{P}_i, & i =1  \\  \mathcal{P}_i - \mathcal{R}'_{i}{M}_{i-1}^{-1}\mathcal{R}_{i}, & i \in \mathbb{N}\backslash\{1\}.
        \end{cases}
        \end{aligned}
    \end{equation}
\end{lemma}

With these results, we now state the conditions for distributed verification of the passivity of the interconnected system $\Sigma$.
\begin{theorem} \label{thm:seq_analysis}
The cascade interconnection \eqref{interconnected_linear} is SP from $w$ to $y$ if
there exist scalars $\epsilon_i$ and matrices $Q_i\in \mathbb{R}^{n_i \times n_i}$, $i\in\mathbb{N}_N$, such that 
\begin{equation}\label{seq_Cholesky}
\begin{aligned}
    \mathbf{P_1}:   \mbox{Find } \quad &\epsilon_i>0, \, Q_i>0 \\
     \mbox{s.t.} \quad & C'_i  = Q_i B_i^{(2)} , 
        \quad \mathcal{M}_i  > 0,
\end{aligned}
\end{equation}
is feasible $\forall i\! \in\!\mathbb{N}_N$, where $\mathcal{M}_i$ is computed in \eqref{messenger_matrix}.
\end{theorem}
\setcounter{equation}{7}
\begin{remark}[Messenger matrix]\label{rem:messengermatrix}
Theorem \ref{thm:seq_analysis} presents a sequential approach to analyze the passivity of a large-scale cascade interconnection of linear dynamical systems. In this approach, each subsystem $\Sigma_i$ computes and stores an information matrix called the \textbf{\textit{messenger matrix}} $\mathcal{M}_i$,~$i~\in~\mathbb{N}_N$. Each messenger matrix $\mathcal{M}_i$ is divided into two parts, (i) $\mathcal{S}_i$, which contains information about the dynamics of the subsystem itself, and (ii) $\mathcal{F}_i$, which contains information about the dynamical coupling with its neighbor $\Sigma_{i-1}$. Computation of the latter uses messenger matrix $\mathcal{M}_{i-1}$ communicated by $\Sigma_{i-1}$ and the coupling terms $\hat{H}_{i-1,i}$ and $\hat{H}_{i,i-1}$ pertaining to the interconnection with $\Sigma_{i-1}$. Since computation of $\mathcal{M}_i$ only requires information about dynamics of subsystem $\Sigma_i$ and limited information from its preceding neighbor, the passivity analysis approach described in Theorem \ref{thm:seq_analysis} is sequential. 
\end{remark}
\begin{figure}[b]
    \centering
    \vspace{-1em}
    \includegraphics[scale=0.5,trim=0.3cm 0cm 0cm 0cm]{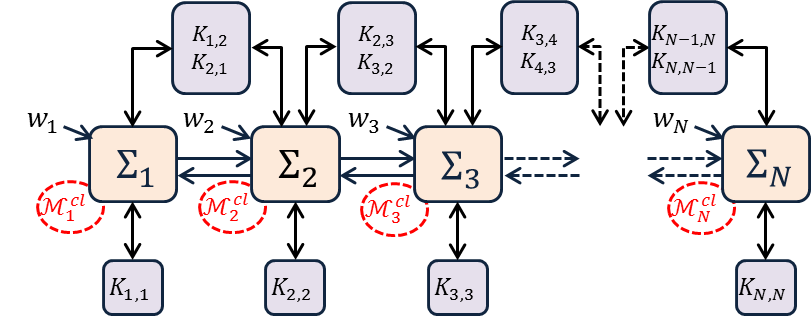}
    \caption{Distributed control architecture for cascade interconnected system.}
    \label{fig:seq_control}
   \vspace{-1.5em}
\end{figure}

\subsection{Sequential Control Design:}\label{subsec:control}
\begin{figure*}[!b]\small
\hrulefill
    \setcounter{MYtempeqncnt}{\value{equation}}
    \setcounter{equation}{10}
        \vspace{0.5em}
    \begin{equation}\label{close_loop_messenger_matrix}
    \begin{aligned} 
    \mathcal{M}^{cl}_i &= \begin{cases} \mathcal{S}_1, & i =1  \\  \mathcal{S}_i - \mathcal{F}_i, & i \in \mathbb{N}_N \backslash \{1\}
        \end{cases} \\
    \hat{H}_{i,j} &= Q_i (B_i^{(1)}H_{i,j}+B_i^{(3)}K_{i,j}), \hspace{11.2em}  i\in\mathbb{N}_N,\, j \in \mathcal{I}_i \\
    \mathcal{S}_i & = -(A'_i Q_i+Q_i A_i)-(\hat{H}_{i,i}+\hat{H}_{i,i}') - \epsilon_i \mbox{I}, \hspace{5.4em} i \in \mathbb{N}_N
    \\\mathcal{F}_i &= (\hat{H}'_{i-1,i}+\hat{H}_{i,i-1})(\mathcal{M}^{cl}_{i-1})^{-1}(\hat{H}_{i-1,i}+\hat{H}'_{i,i-1}), \hspace{2.2em} i \in \mathbb{N}_N\backslash \{1\}
    \end{aligned}
    \end{equation}\normalsize
\vspace{-0.6em}
\setcounter{equation}{\value{MYtempeqncnt}}
\end{figure*}
\setcounter{equation}{7}
In Theorem \ref{thm:seq_analysis}, we presented a set of conditions sufficient to guarantee passivity of \eqref{interconnected_linear}. 
We now describe a control design algorithm that utilizes Theorem \ref{thm:seq_analysis} to synthesize subsystem-level passivating state feedback controllers in a distributed manner. The controllers are designed at every subsystem locally, that is, without using any direct information about the dynamics of other subsystems in the interconnection. We assume that limited information in the form of the \textit{messenger matrix} as discussed in Remark \ref{rem:messengermatrix} can be exchanged between a subsystem and its immediate neighbor. The local control law for every subsystem $\Sigma_i$, $i\in\mathbb{N}_N$ is given by
\vspace{-0.2em}
\begin{equation}\label{control_law}
         u_i = \sum\limits_{j\in\mathcal{I}_i}K_{i,j}x_j, \vspace{-0.4em}
\end{equation}
where $K_{i,j}$ are the local state feedback gain matrices for the subsystem pair $(\Sigma_i,\Sigma_j), i \in \mathbb{N}_N, j \in \mathcal{I}_i$. Then, the dynamics of the closed loop cascade interconnected system with the control law \eqref{control_law} is given by
\vspace{-0.3em}
    \begin{equation}\label{interconnected_linear_closed}
    \begin{aligned}
	\dot{x}(t)  &= (A+B^{(1)}H+B^{(3)}K)x(t) + B^{(2)}w(t)\\
	y(t)  &= Cx(t) \\
	K & = \left[K_{i,j}\right]_{i,j\in\mathbb{N_N}}, \quad K_{i,j} = 0 \ \forall \vert i-j\vert>1.  \vspace{-1em}
\end{aligned}
\end{equation}
The following corollary presents an approach to design these subsystem-level controllers in a sequential manner, starting at one end of the cascade interconnection. 

\begin{corollary} \label{cor:seq_design}
The closed loop interconnection \eqref{interconnected_linear_closed} 
is SP if
\begin{equation}\label{close_loop_seq_Cholesky}
\begin{aligned}
    \mathbf{P_2}:   \mbox{Find } \quad &\epsilon_i>0, \, Q_i>0, \, K_{i,j}, \, j \in \mathcal{I}_i  \\
     \mbox{s.t.} \quad & C'_i  = Q_i B_i^{(2)}, 
        \quad \mathcal{M}^{cl}_i  > 0, \\
        & \epsilon_i \in \mathbb{R}, \, Q_i \in \mathbb{R}^{n_i \times n_i}, \, K_{i,j} \in \mathbb{R}^{p_i \times n_j}, 
\end{aligned}
\end{equation}
is feasible $\forall i\in\!\mathbb{N}_N$, where $\mathcal{M}^{cl}_i$ is the closed-loop messenger matrix of $\Sigma_i$ computed from \eqref{close_loop_messenger_matrix}.
\end{corollary}
\setcounter{equation}{11}

The proof follows from using \eqref{interconnected_linear_closed} in Theorem~\ref{thm:seq_analysis}.

\begin{figure}[b]
    \centering
   \vspace{1em}
    \includegraphics[scale=0.48,trim=0cm 0.9cm 0cm 0.5cm]{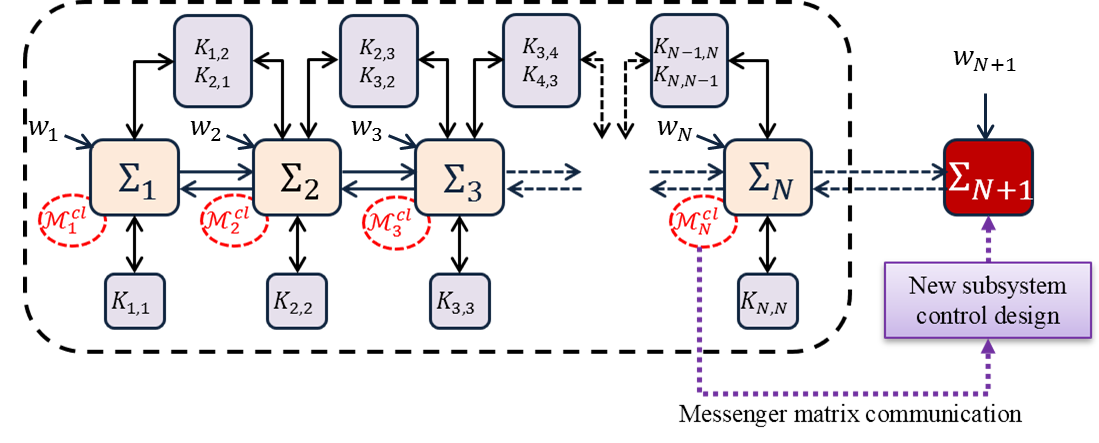}
    \caption{Schematic of compositional control design.}
    \label{fig:compositional_control}
   \vspace{-1em}
\end{figure}

The distributed control architecture for the cascade interconnection is shown in Fig. \ref{fig:seq_control}. The distributed controller for each subsystem $\Sigma_i$, $i \in \mathbb{N}_N$ is composed of two terms - (i) self state feedback gain $K_{i,i}$, and (ii) coupling state feedback gain $K_{i,j}$, $j\in\mathcal{I}_i$. Therefore, every subsystem uses its own state and the state of its immediate neighbors to compute its control action. Every subsystem also stores its closed loop messenger matrix $M^{cl}_i$, which is communicated to its immediate neighbors during the sequential control design process as described in Algorithm 1. Note that $\mathbf{P}_2$ may not always be feasible - for example, in the case of the linear subsystems being nonminimum phase.

\begin{algorithm}[H]
	\caption{Sequential control design}
	\label{algo_seq}
	\begin{algorithmic}[1]
	    \State Initialize $i = 1$, $\mathcal{M}^{cl}_0 = 0$. 
	    \While{$i\leq N$, at subsystem $\Sigma_i$,} \label{loop}
	    \State Receive $\mathcal{M}^{cl}_{i-1}$ and $P_{i-1}$ from $\Sigma_{i-1}$.
	    \State Set $\mathcal{M}_{i-1} = \mathcal{M}^{cl}_{i-1}$.
	    \If{$\mathbf{P_1}$ is feasible}
	    \State Compute $\mathcal{M}_i>0$ from \eqref{seq_Cholesky} and \eqref{messenger_matrix} and set $\mathcal{M}^{cl}_i = \mathcal{M}_i$.
	    \State Set $K_{i,i}=K_{i,i-1}=K_{i-1,i}=0$.
	    \Else
	    \State{\textit{Control design:}} Solve $\mathbf{P_2}$ to compute $K_{i,i}$, $K_{i,i-1}$, $K_{i-1,i}$ and $\mathcal{M}^{cl}_i>0$ from  \eqref{close_loop_seq_Cholesky} and \eqref{close_loop_messenger_matrix}.
	    \EndIf
	    \State Send $K_{i-1,i}$ to $\Sigma_{i-1}$.
		\State Set $i\mapsto i+1$.
		\EndWhile
	\end{algorithmic}
\end{algorithm}
\vspace{-1.1em}

\subsection{Compositionality}\label{subsec:comp}

In large-scale networks where new subsystems may be added to the pre-existing interconnection at a later time, it is desirable to develop a framework which facilitates addition of new subsystems without having to re-design the pre-existing controllers. This property is referred to as \textit{compositionality}. Fig. \ref{fig:compositional_control} shows the schematic for compositional implementation of the proposed control design. When a new subsystem $\Sigma_{N+1}$ is added to $\Sigma$, it receives the closed loop messenger matrix from its neighbor $\Sigma_N$. Then, the design procedure in Algorithm \ref{algo_seq} can be used just for $\Sigma_{N+1}$ to obtain its closed loop messenger matrix $\mathcal{M}^{cl}_{N+1}$, self feedback controller gain $K_{N+1,N+1}$ and coupling feedback controller gains $K_{N+1,N}$ and $K_{N,N+1}$ corresponding to the new interconnection between $\Sigma_{N}$ and $\Sigma_{N+1}$. This procedure is compositional since there is no limit to the number of subsystems that can be added in this manner. 

\section{Example}\label{sec:Example} 
\begin{figure}[!t]
    \centering
    \vspace{0.8em}
    \includegraphics[scale=0.65, trim = 0cm 0cm 0cm 0.5cm]{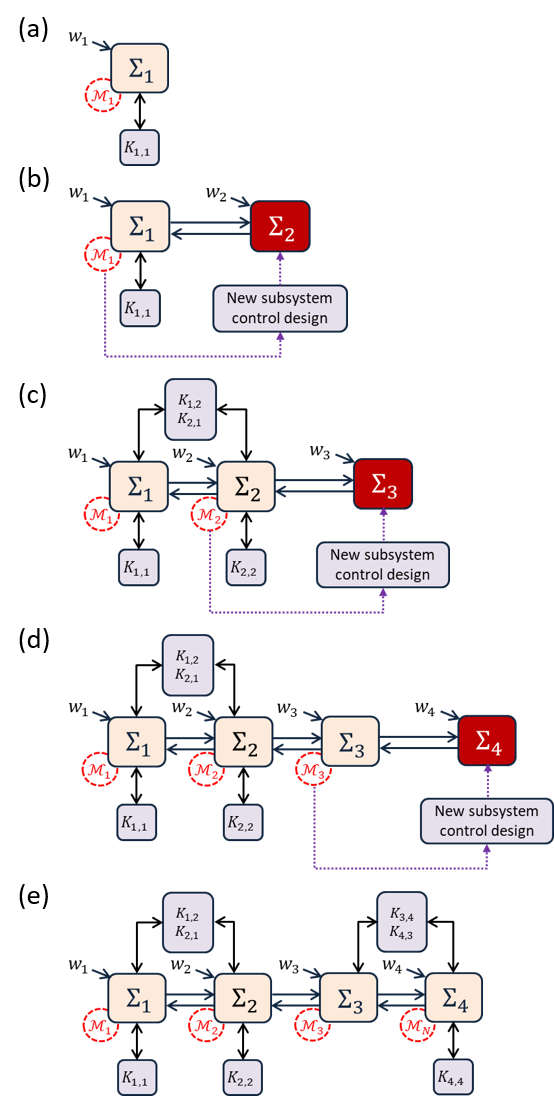}
    \caption{Compositional control synthesis for example.}
    \vspace{-1.5em}
   \label{fig:example}
\end{figure}

\setcounter{equation}{11}
In this section, we present an example to demonstrate the applicability of our results in compositional control synthesis for dynamically growing interconnections, where new subsystems are added to the cascade after the controllers are designed for the existing system. We start by considering a simple second order system $\Sigma_1$ with dynamics \small
\begin{align}
    \dot{x}_1(t) & = \begin{bmatrix} -9 & 1\\ 5 & 7\end{bmatrix} x_1 (t)\! +\! \begin{bmatrix} 1 \\ 1 \end{bmatrix} v_1(t)\! +\! \begin{bmatrix} 1 \\ 0.5 \end{bmatrix} w_1(t)\! +\! \begin{bmatrix} 1 \\ 1 \end{bmatrix} u_1(t) \nonumber \\
    y_1(t) & = \begin{bmatrix} 3 & 2 \end{bmatrix} x_1(t) \\
    v_1(t) & = \begin{bmatrix} 0.5 & -0.7 \end{bmatrix} x_1(t) \nonumber.
\end{align}\normalsize
The objective is to guarantee passivity of system $\Sigma_1$. We use Algorithm 1 to check the sufficient conditions in Theorem \ref{thm:seq_analysis} and compute an appropriate passivating controller gain matrix $K_{1,1}$ and closed loop messenger matrix $\mathcal{M}_1$ as shown in Fig. \ref{fig:example}~(a). Now suppose, a new subsystem $\Sigma_2$, \small
\begin{align}
    \dot{x}_2(t) & = 3 x_2 (t)\! +\! v_2(t)\! +\!  w_2(t)\! +\! u_2(t) \nonumber \\
    y_2(t) & = x_2(t) \\
    v_2(t) & = \begin{bmatrix} 1 & -0.5 \end{bmatrix} x_1(t) +  0.5 x_2(t) \nonumber,
\end{align} \normalsize
is added in cascade to $\Sigma_1$, as shown in Fig. \ref{fig:example}~(b). With new coupling between $\Sigma_1$ with $\Sigma_2$, the coupling input $v_1(t)$ is updated to 
$v_1(t)  = \begin{bmatrix} 0.5 & -0.7 \end{bmatrix} x_1(t) + 0.1 x_2(t).$ \normalsize

In order to verify if this interconnection of $\Sigma_1$ and $\Sigma_2$ is passive, as per the sequential verification algorithm, the messenger matrix $\mathcal{M}_1$ is transmitted to $\Sigma_2$. Since $\mathbf{P}_1$ is not feasible for this added subsystem, passivating control synthesis is carried out for subsystem $\Sigma_2$ and its interconnection with $\Sigma_1$. Corollary \ref{cor:seq_design} is used to design subsystem-level controller gains $K_{2,2}$, $K_{1,2}$ and $K_{2,1}$. The messenger matrix $\mathcal{M}_2$ pertaining to subsystem $\Sigma_2$ is also computed.

We now add a third subsystem $\Sigma_3$ to this already existing cascade interconnection of subsystems $\Sigma_1$ and $\Sigma_2$. The new subsystem $\Sigma_3$ receives messenger matrix $\mathcal{M}_2$ from $\Sigma_2$ (Fig. \ref{fig:example}~(c)). The dynamics of $\Sigma_3$ is given by \small
\begin{align}
    \dot{x}_3(t) & = - x_3 (t)\! +\!  v_3(t)\! +\!  w_3(t)\! +\!  u_3(t) \nonumber \\
    y_3(t) & = x_3(t) \\
    v_3(t) & = -0.7 x_2(t) + 0.2 x_3(t) \nonumber.
\end{align}\normalsize
The new coupling input $v_2(t)$ for subsystem $\Sigma_2$ is, 
$v_2(t)  = \begin{bmatrix} 1 & -0.5 \end{bmatrix} x_1(t) +  0.5 x_2(t) -0.1 x_3(t)$. This time, when problem $\mathbf{P}_1$ is solved for $i=3$, it is feasible. Therefore, the messenger matrix $\mathcal{M}_3$ is computed but no controllers are designed. Similarly, we add a fourth subsystem $\Sigma_4$, \small
\begin{align}
    \dot{x}_4(t) & = \begin{bmatrix} 2 & 1\\ 3 & 0.8\end{bmatrix} x_4(t)\! +\! \begin{bmatrix} 1.2 \\ 0.8 \end{bmatrix} v_1(t)\! +\! \begin{bmatrix} 0.5 \\ -0.2 \end{bmatrix} w_4(t)\! +\! \begin{bmatrix} 1.2 \\ 0.8 \end{bmatrix} u_4(t) \nonumber \\
    y_4(t) & = \begin{bmatrix} 2.1 & 0.6 \end{bmatrix} x_4(t) \\
   v_4(t) & = -0.9 x_3(t)+\begin{bmatrix} 1.1 & 0.4 \end{bmatrix} x_4(t) \nonumber,
\end{align} \normalsize
to this cascade interconnection, as shown in Fig. \ref{fig:example}~(d). The new coupling input $v_3(t)$ is,
$v_3(t) = -0.7 x_2(t) + 0.2 x_3(t) + \begin{bmatrix}0.2 & 0.2 \end{bmatrix}x_4(t).$\normalsize

\begin{figure}[t]
\vspace{0.5em}
    \centering
    \includegraphics[scale=0.65, trim = 0cm 0.19cm 0cm 2.5cm, clip]{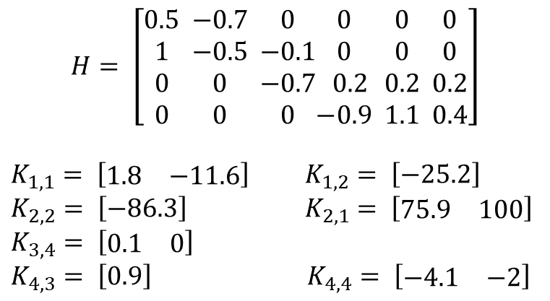}
    \caption{Coupling matrix $H$ for cascade interconnection of subsystems $\Sigma_1$, $\Sigma_2$, $\Sigma_3$ and $\Sigma_4$, and controller gains for the subsystem-level controllers.}
   \label{fig:example-control}
  \vspace{-0.8em}
\end{figure}
Using the messenger matrix $\mathcal{M}_3$ received from $\Sigma_3$, we can complete the passivity verification and design process for this new subsystem. The designed controllers corresponding to the coupling between $\Sigma_3$ and $\Sigma_4$ guarantee that the entire cascade interconnection including $\Sigma_4$ is SP. We can thus continue to add subsystems to the interconnection in a compositional manner. This step-by-step sequential verification and control synthesis process is outlined in Fig. \ref{fig:example}. We make the following remarks:

\begin{enumerate}
    \item The sequential verification and controller synthesis algorithm does not require that all the subsystems be of the same order, as long as the input-output dimensions are suitable to define the  interconnection.
    \item For compositional synthesis, the complexity of proposed algorithm does not depend on the number of subsystems already existing in the interconnection.
    \item While the control synthesis equations \eqref{close_loop_seq_Cholesky}-\eqref{close_loop_messenger_matrix} in $\mathbf{P}_2$ are nonlinear matrix inequalities, it is fairly straightforward to express them as linear matrix inequalities using a Schur's complement method \cite[Section 4.6]{vanantwerp2000tutorial}.
\end{enumerate}

\section{Conclusion}
We presented a sequential control design framework for passivation of cascade interconnected systems, where both the control law and the control design process are distributed, requiring only local information. 
A similar procedure can also be used for distributed synthesis of local stabilizing controllers as opposed to passivating controllers. 

\appendix

\noindent \textbf{Proof of Lemma \ref{lemma:Cholesky_tri_diagonal}}
\newline
Consider lower triangular matrices $\mathcal{U}_i, \ i\in \mathbb{N}_N$ and matrix 
\footnotesize
\begin{align}\label{cholesky_decomp_lem}
    L&= \begin{bmatrix} 
    \diagentry{\mathcal{U}_1} \\
    \diagentry{\mathcal{V}_2} & \diagentry{\mathcal{U}_2} &  & \text{\huge 0}\\
    &\diagentry{\mathcal{V}_3} & \diagentry{\mathcal{U}_3}\\
    &&\diagentry{\xddots}& \diagentry{\xddots} \\
    &\text{\huge 0}&& \diagentry{\mathcal{V}_{N-1}} & \diagentry{\mathcal{U}_{N-1}} \\
    &&&&\diagentry{\mathcal{V}_N} & \diagentry{\mathcal{U}_N}\\
    \end{bmatrix},\end{align}
    
    \begin{align}
        \mathcal{U}_i\mathcal{U}'_i \!&=\! \begin{cases} \!\mathcal{P}_i & \mbox{if } i =1\\
        \!\mathcal{P}_i \!-\! \mathcal{R}'_{i}(\mathcal{U}_{i-1}\mathcal{U}'_{i-1})^{-1}\mathcal{R}_{i} & \forall i \in \{2,\ldots,N\} \end{cases} \\
        \mathcal{V}_i \!&=\! \mathcal{R}'_i(\mathcal{U}'_{i-1})^{-1}, \hspace{6.3em} \forall i \in \{2,\ldots,N\},
    \end{align}
\normalsize with $\mathcal{P}_i$ and $\mathcal{R}_i$ being the elements of $P$ as defined in \eqref{tri_diag_matrix}. Define ${M}_i = \mathcal{U}_i\mathcal{U}'_i$, $\forall i\in\mathbb{N}_N$. From Lemma \ref{lemma:Cholesky}, if \eqref{seq_Cholesky_lemma} holds, $\mathcal{U}_i$ will exist with positive diagonal entries. The invertibility of $\mathcal{U}_i, \, i\in\mathbb{N}_N$ guarantees the existence of $\mathcal{V}_j, \, j\in\mathbb{N}_N$. Therefore, we can always find a lower triangular matrix $L$ of the form \eqref{cholesky_decomp_lem}, with positive diagonal entries, such that $P = LL'$. From Lemma \ref{lemma:Cholesky}, $P>0$. This proves the sufficiency of Lemma \ref{lemma:Cholesky_tri_diagonal}.  Along similar lines, we can also prove the necessity of \eqref{seq_Cholesky} for positive definiteness of $P$. $\blacksquare$ 


\vspace{0.7em}
\noindent \textbf{Proof of Theorem \ref{thm:seq_analysis}}

\noindent Consider a positive function $V(x(t)) = \frac{1}{2}x'Qx$, $\quad Q>0$. For the cascade interconnected linear system \eqref{interconnected_linear}, with $u\equiv 0$, \footnotesize
 \begin{align}
     w'y -  \epsilon x'x - \dot{V} & = \frac{1}{2}w'y + \frac{1}{2}y'w -  \epsilon x'x- \frac{1}{2}(\dot{x}'Qx + x'Q\dot{x})  \nonumber\\
     & = - \frac{1}{2}\begin{bmatrix}x \\ w\end{bmatrix}' \Gamma \begin{bmatrix}x \\ w\end{bmatrix},
 \end{align}
 $$\Gamma \! = \! \begin{bmatrix}
-(A + B^{(1)}H)'Q-Q(A + B^{(1)}H)-\epsilon \mbox{I} & -QB^{(2)}\!+\!C' \\
     -B^{(2)'}Q+C & 0
     \end{bmatrix}. $$ \normalsize If $\Gamma \geq 0$, \eqref{interconnected_linear} is SP. Taking the Schur's complement of $\Gamma$, \footnotesize
     \begin{equation}\label{SP_cond_proof}
     \begin{aligned}
         &\mathbf{W} = -(A + B^{(1)}H)'Q-Q(A + B^{(1)}H) - \epsilon \mbox{I}\geq 0\\
         &Q_i B_i^{(2)}  = C_i', \quad Q_i>0 \quad \forall i \in \mathbb{N}_N \\
         & Q = \mbox{diag}(Q_1, Q_2, \ldots, Q_N).
     \end{aligned}
     \end{equation} \normalsize
For the interconnected linear system \eqref{interconnected_linear}, the sparsity of the interconnection matrix $H$ dictates the structure of the matrix $\mathbf{W}$. For cascade interconnections, $\mathbf{W}$ is block tri-diagonal, with \footnotesize
\begin{subequations}
    \begin{align}
        \mathbf{W} & = \begin{bmatrix} 
        \diagentry{\mathcal{P}_1} & \diagentry{\mathcal{R}_2} \\
        \diagentry{\mathcal{R}'_2} & \diagentry{\mathcal{P}_2} & \diagentry{\mathcal{R}_3} & & \text{\huge 0}\\
        &\diagentry{\mathcal{R}'_3} & \diagentry{\mathcal{P}_3} & \diagentry{\mathcal{R}_4}\\
        &&\diagentry{\xddots}& \diagentry{\xddots} & \diagentry{\xddots}\\
        &\text{\huge 0}&& \diagentry{\mathcal{R}'_{N-1}} & \diagentry{\mathcal{P}_{N-1}} & \diagentry{\mathcal{R}_N}\\
        &&&&\diagentry{\mathcal{R}'_N} & \diagentry{\mathcal{P}_N}\\
        \end{bmatrix}, \\
        \mathcal{P}_i &= -(A'_i Q_i+Q_i A_i)-(\hat{H}_{i,i}+\hat{H}'_{i,i}) - \epsilon\mbox{I}, \hspace{0.8em} i \in \mathbb{N}_N, \\
        \mathcal{R}_j &= -(\hat{H}_{j-1,j}+\hat{H}'_{j,j-1}), \hspace{6em} j \in \{2, \ldots, N\}.
    \end{align} 
\end{subequations} \normalsize
If \eqref{seq_Cholesky} and \eqref{messenger_matrix} hold, then, from Lemma \ref{lemma:Cholesky_tri_diagonal}, $\mathbf{W}>0$ and all conditions in \eqref{SP_cond_proof} are satisfied with $\epsilon = \mbox{min}(\epsilon_1,\epsilon_2, \ldots, \epsilon_N)$. Therefore, linear system \eqref{interconnected_linear} is SP. $\blacksquare$   

\balance
\bibliographystyle{IEEEtran}
\bibliography{references}

\end{document}